\documentclass[prd,a4paper,showpacs,byrevtex]{revtex4}

\usepackage{graphicx}
\usepackage{dcolumn}
\usepackage{amsmath}
\usepackage{array}
\usepackage{bm}
\usepackage{amssymb}
\usepackage{amsfonts}
\usepackage{epic} 
\usepackage{calc} 
\usepackage[noautoscale,vcentermath]{youngtab}

\begin{document}


\title{Constituent gluon interpretation of glueballs and gluelumps}

\author{Nicolas \surname{Boulanger $^a$}}
\thanks{``Progetto Italia'' post-doctoral fellow}
\email[E-mail: ]{nicolas.boulanger@sns.it}
\author{Fabien \surname{Buisseret $^b$}}
\thanks{FNRS Research Fellow}
\email[E-mail: ]{fabien.buisseret@umh.ac.be}
\author{Vincent \surname{Mathieu $^b$}}
\thanks{IISN Scientific Research Worker}
\email[E-mail: ]{vincent.mathieu@umh.ac.be}
\author{Claude \surname{Semay $^b$}}
\thanks{FNRS Research Associate}
\email[E-mail: ]{claude.semay@umh.ac.be}
\affiliation{$^a$ Scuola Normale Superiore,
Piazza dei Cavalieri 7, 56126 Pisa, Italy\\
$^b$ Groupe de Physique Nucl\'{e}aire Th\'{e}orique,
Universit\'{e} de Mons-Hainaut,
Acad\'{e}mie universitaire Wallonie-Bruxelles,
Place du Parc 20, BE-7000 Mons, Belgium}

\date{\today}

\begin{abstract}
Arguments are given that support the interpretation of the lattice QCD glueball and gluelump spectra in terms of bound states of massless constituent gluons with helicity-1. In this scheme, the mass hierarchy of the currently known gluelumps and glueballs is mainly due to the number of constituent gluons and can be understood within a simple flux tube model. It is also argued that the lattice QCD $0^{+-}$ glueball should be seen as a four-gluon bound state. 
\end{abstract}


\maketitle

\section{Introduction}

\begin{figure}[ht]
\includegraphics[width=9.0cm]{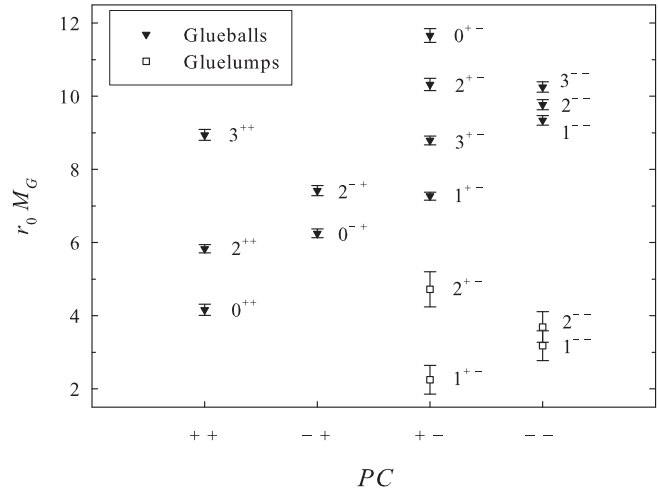}
\caption{Plot of the lattice QCD glueball spectrum (triangles). Data are taken from Ref.~\cite{lat3}. The four lowest-lying gluelumps are also plotted (squares). Gluelump masses are taken from Ref.~\cite{gl1}. Masses are given in units of the lattice energy scale $r_0^{-1}$. }
\label{fig1}
\end{figure}

QCD allows the existence of purely gluonic bound states, called glueballs, whose structure and properties deserve much interest (see Ref.~\cite{exp_gg} for a recent review). An important theoretical achievement in this field has been the computation of the glueball spectrum in lattice QCD \cite{lat1,lat2,lat3}. A plot of this spectrum is given in Fig.~\ref{fig1}, where the masses are expressed in units of the lattice energy scale $r^{-1}_0$ in order to avoid larger error bars due to the determination of this energy scale. Notice that $r^{-2}_0$ can be understood as the energy density of a fundamental flux tube, that is, the flux tube linking a quark-antiquark pair. Apart from lattice QCD, the glueball spectrum has also been computed by using effective approaches like Coulomb gauge QCD \cite{cg} and potential models (see for example Refs.~\cite{bar,bar2,brau,gvm,gvm2}). In potential models, glueballs are assumed to be bound states of constituent gluons. Within this framework, gluons are supposed to be massless or not, with either a helicity-1 or a spin-1. The physical properties of constituent gluons are actually still a matter of controversy, and even the constituent gluon picture could be questioned since potential models have serious difficulties in reproducing all the currently known lattice QCD data. It has been shown in Ref.~\cite{gluh1} that the lightest glueballs, which have $C=+$, can be successfully modeled by a two-gluon system in which the constituent gluons are massless helicity-1 particles. The proper inclusion of the helicity degrees of freedom dramatically improves the compatibility between lattice QCD and potential models \cite{gluh1}. This picture is also supported by the Coulomb gauge study of Ref.~\cite{cg}.

Motivated by these previous results, we propose in the present paper an interpretation of the whole glueball spectrum from a helicity-1 constituent gluon point of view. The relevance of this picture and its links to QCD are discussed in Sec.~\ref{motiv}, and the general principles of our classification scheme are exposed in Sec.~\ref{gene}. Glueballs and gluelumps are then analyzed in Secs.~\ref{sec2} and \ref{sec3} respectively. Gluelumps are actually bound states of the gluonic field in a static color octet source that have been studied first in lattice QCD \cite{gl2}. Only a relative mass spectrum was computed in this last reference, while a determination of the absolute mass spectrum has been given in Ref.~\cite{gl1}. The lowest-lying gluelump states are also plotted in Fig.~\ref{fig1}. What we argue is that the low-lying gluelumps can be interpreted as bound states of a single constituent gluon. A simple flux tube potential model is then proposed in Sec.~\ref{sec4} in order to understand the mass hierarchy of gluelumps and glueballs. By combining the flux tube model and the helicity formalism, we will finally compute in Sec.~\ref{4glu} the mass of the lowest-lying $0^{+-}$ four-gluon state and compare it to the lattice QCD data. Conclusions will then be drawn in Sec.~\ref{conc}. Appendix~\ref{hsgl} gives some two-body helicity states not built in Ref.~\cite{gluh1}, while Appendix~\ref{cval} recalls the way of computing the $C$-parity of few-gluon glueballs.

\section{Constituent gluons and QCD}\label{motiv}
\subsection{Effective approaches}

The basic assumption underlying our work is that a description of gluelumps and glueballs in
terms of states with a given number of constituent gluons is relevant, or at least that it can
be a satisfactory approximation of what these gluonic hadrons exactly are. The idea that
low-energy QCD allows for an effective description of hadrons as bound states of constituent
particles is not new: The classification of baryons and mesons with the quark hypothesis is a
first historical example of the viability of such a picture. However, there should exist a way
to connect the constituent approaches to QCD. To our knowledge, this longstanding problem is
far from being completely solved, but interesting results have been obtained within different
frameworks. We recall some of these approaches and comments on them in the following.

First, we turn our attention to large-$N_c$ SU($N_c$)-gauge theories, where $N_c$ is the
number of colors. The $1/N_c$ expansion in QCD, valid in particular at low energy, has
revealed itself to be a powerful tool which has played an important role in numerous
theoretical aspects of QCD (see e.g. Ref.~\cite{mata} for useful references). In the limit
where the number of colors becomes very large, it has been shown that the description of
baryons, made of $N_c$ valence quarks, becomes simpler \cite{witten}. Model independent mass
formulae can be obtained within this framework and lead to a very accurate description of
experimentally known baryons, either light or heavy \cite{mata}; thus the large-$N_c$ limit
appears to capture the essential features of QCD, which is characterized by ``only" $N_c=3$.
Moreover, it has recently been shown that the mass formulae obtained from a constituent quark
model are in remarkable agreement with those arising from the large-$N_c$ expansion in the
light baryon sector \cite{lnc}. This comforts the constituent quark picture in the most
difficult cases, when light (even massless) quarks are present. Of importance for our purpose
is the recent observation that the number of valence gluons is a good quantum number for
glueballs in the large-$N_c$ limit \cite{liu}. By analogy with the baryons, this suggests that
the Fock space expansion of glueballs at $N_c=3$ is dominated by a particular component,
characterized by its number of constituent gluons. In this spirit, the present work could be
seen as an attempt to identify these dominant components for each known glueball and gluelump.

Second, we can get useful informations from Coulomb gauge QCD \cite{cg0}. In this approach,
the QCD Hamiltonian is written with the gluonic field in the Coulomb gauge. A key feature of
this gauge is that the elimination of the nondynamical degrees of freedom creates an
instantaneous nonperturbative interaction. Properties of low-energy hadronic states are
eventually computed from this Hamiltonian and the Fock space in which these states are defined
is expanded on a quasiparticle basis, either quarks or gluons. A general property of hadronic
states is then that their Fock space expansion converges quickly. This leads to a connection
with constituent models since, again, it seems relevant to assume that a fixed given number of
valence quarks or gluons can be associated with a given hadron. The notion of an instantaneous
confining potential also naturally emerges from Coulomb gauge QCD. The connections between
constituent quark models and Coulomb gauge QCD have been discussed in Ref.~\cite{cg1} while
glueballs are studied in Refs.~\cite{cg2}. It is worth mentioning that in these last works,
the low-lying glueballs are successfully described in terms of two-transverse gluon bound
states, with a negligible three-gluon component. These conclusions are shared by the recent
results of Ref.~\cite{gluh1}, where pure two-transverse gluon bound states have been shown to
accurately reproduce the lattice QCD and Coulomb gauge QCD data.

In Coulomb gauge QCD as in many potential models, the constituent particles have a particular
status since they are confined. In particular they gain a constituent mass, different from
their bare mass, because of the confining interaction. One can look in particular at the
Coulomb gauge study of Ref.~\cite{llan1}, where it is shown that constituent gluons, although
having a zero bare mass, gain a running constituent mass which is about $0.7$~GeV at zero
momentum. They become then distinguishable from the exchanged gluons, which mediate the
interaction without being themselves confined.

An effective QCD Hamiltonian for various hadronic systems can also be derived from the field
correlator method \cite{first}. In these models, the picture of constituent particles (quarks
or gluons) linked by color flux tubes emerges naturally. For instance, the constituent mass of
the confined gluons is defined as $\left\langle \vec p^{\, 2}\right\rangle$, that is their
average kinetic energy. Interestingly, this quantity is also typically of about 0.7~GeV for
the glueball ground state.

\subsection{The correlator method }

We have just shown that several effective approaches of low-energy QCD agree with the constituent gluon picture for glueballs. Even if the Fock space expansion of a glueball or gluelump may be nontrivial, these models actually suggest that one component with a given number of valence gluons may be dominant. Now we make some comments on the correlator method, that allows to compute glueball masses in lattice QCD for example \cite{glulat}.

Let us first consider the initial state (at zero euclidian time) $\left|\Psi(0)\right\rangle=\Phi^{(R)}(0)\left|0\right\rangle$, where $\Phi^{(R)}(0)$ is an operator creating a glueball or a gluelump out of the QCD vacuum $\left|0\right\rangle$, and where $R$ denotes the irrep corresponding to particular $J^{PC}$ quantum numbers. Although highly nontrivial, the evolution of this state is computable, with lattice methods in particular \cite{glulat,book}. After a time $\tau$ one gets the state $\left|\Psi(\tau)\right\rangle=\Phi^{(R)}(\tau)\left|0\right\rangle$, and the correlator $C(\tau)=\left\langle\Psi(\tau) \right|\left.\Psi(0)\right\rangle$ can be evaluated. As intuitively expected, $\lim_{\tau\rightarrow\infty}C(\tau)\propto {\rm e}^{-m_G\, \tau}$, $m_G$ being the mass of the lightest state which can be created by the operator $\Phi^{(R)}$. Examples of such operators are given in Table~\ref{tabop}. A rule giving the mass hierarchy of glueballs has been formulated in Ref.~\cite{jaffe}: The higher the mass dimension of an operator is, the heavier the resulting hadron is. For example, the lowest $1^{+-}$ gluelump operator has dimension $2$, while the first $0^{++}$ and $1^{+-}$ glueballs have dimension 4 and 6 respectively. Figure~\ref{fig1} shows that the observed lattice QCD mass hierarchy follows this dimension rule. There are two ways of increasing the dimension of an operator: Either to increase the order of the operator in the chromo-electric or -magnetic fields $\vec E_a$ and $\vec B_a$ respectively, or to add a covariant derivative $D^i$ (as in the $2^{--}$ gluelump). Both ways have a very natural explanation within a constituent gluon picture. Indeed, a transverse gluon creator is contained in each $\vec E_a$ or $\vec B_a$ factor. Assuming thus that each such factor generates a constituent gluon, we find that gluelumps are made of a single constituent gluon, and that they are logically lighter than glueballs, which are made of at least two constituent gluons. 
Another illustration of this interpretation is that a $C$-odd glueball can be created with at least a cubic expression in the fields: As possible three-gluon states, they logically appear to be heavier than two-gluon states. Within this picture, the addition of totally symmetrized covariant derivatives does not affect the number of constituent gluons, but adds momentum: The covariant derivative rather generates excitations of the considered few-gluon system \cite{jaffe}.

Even if $\left|\Psi(0)\right\rangle$ can be seen as a state with a given number of constituent gluons, because it is created by operators such as those of Table~\ref{tabop}, the final physical state $\left|\Psi(\tau\rightarrow\infty)\right\rangle$ could have a far more complicated Fock space expansion due to the complex QCD interactions that come into play. But, as mentioned above, this does not seem to be the case. At this stage a connection with potential models can be made: They all rely on the assumption that it is possible to find an appropriate effective potential satisfactory reproducing the properties of physical hadrons ($\approx\left|\Psi(\tau\rightarrow\infty)\right\rangle$) seen as systems of constituent particles ($\approx\left|\Psi(0)\right\rangle$) interacting \textit{via} this effective potential. To conclude, it is worth mentioning the results of Ref.~\cite{effpot}, where it has been shown that the lattice QCD $0^{++}$ glueball wave function is compatible with the wave function of a flux tube Hamiltonian of the form $2\sqrt{\vec p^{\, 2}}+ar-\kappa/r$ in the $0^{++}$ channel. We recover the idea that the scalar glueball can be seen as a dominantly two-gluon state, where the two constituent gluons ($\approx2\sqrt{\vec p^{\, 2}}$) interact \textit{via} the effective potential $ar-\kappa/r$.

\begin{table}[h]
    \centering
    \caption{Examples of operators creating glueballs and gluelumps with a given $J^{PC}$. The $()$ denote a complete symmetrization. The repeated color indices $a$, $b$, $c$ are intended to be summed and the superscripts $i$, $j$ are space indices.}
    \setlength{\extrarowheight}{3pt}
        \begin{tabular}{clcclcl}
    \hline\hline
    \multicolumn{2}{c}{ Gluelumps \cite{gl1}} && \multicolumn{4}{c}{Glueballs \cite{jaffe}}\\

    $J^{PC}$ & Operator &       &$J^{PC}$ & Operator &  $J^{PC}$ & Operator \\
    \hline
    $1^{+-}$ & $B^i_{a}$ && $0^{++}$ & $\vec E^2_a\pm\vec B^2_a$ & $1^{+-}$ & $d_{abc} (\vec E_a\cdot\vec E_b)\vec B_c$ \\
    $1^{--}$ & $E^i_{a}$ && $0^{-+}$ & $\vec E_a\cdot\vec B_a$ & $1^{--}$ & $d_{abc} (\vec E_a\cdot\vec E_b)\vec E_c$ \\
    $2^{--}$ & $D^{(i}B^{j)}_a$ && $2^{++}$ & $E^i_a E^j_a\pm B^i_aB^j_a-\frac{1}{3}\delta^{ij}(\vec E^2_a\pm\vec B^2_a)$ & $2^{+-}$ & $d_{abc}\left[E^i_a(\vec B_b\times \vec E_c)^j+(i\leftrightarrow j)\right]$ \\
    \hline\hline
        \end{tabular}

    \label{tabop}
\end{table}
\section{General principles}\label{gene}

Following the above discussion, it seems very reasonable to build our study from this
assumption: Glueballs and gluelumps can be seen, at least in first approximation, as bound
states of constituent gluons, these being massless particles with helicity-1. The quantum
states describing bound states of few-gluon systems can be efficiently formulated within the
helicity formalism, introduced in Ref.~\cite{jaco}. Hereafter, we recall the main points of
this formalism. Let us denote $\left|\psi(\vec p,\lambda)\right\rangle$ the quantum state of a
particle with momentum $\vec p$ and helicity $\lambda$. If this particle is massive of spin
$s$, then $\lambda$ can take the $2s+1$ allowed values for a spin projection along a given
axis. But, if the particle is massless, only $\lambda=\pm s$ is allowed. The question is now:
How to write a two-particle helicity state in the rest frame of the system such that both
particles are coupled to a given value of the total spin $\vec J$ with a projection $J_z=M$
and with a given parity? The general answer is given in Ref.~\cite{jaco}, from which it can be
deduced that the helicity states
\begin{eqnarray}\label{hstate}
\left|\lambda_1,\lambda_2;J^P,M,\epsilon\right\rangle&= \frac{1}{\sqrt{2}} &\left\{ \Omega^J_{M,\lambda_1-\lambda_2}\left[\left|\psi(\vec p,\lambda_1)\right\rangle\otimes\left|\psi(-\vec p,\lambda_2)\right\rangle\right]\right.\nonumber\\ && \left. +\epsilon\, \Omega^J_{M,\lambda_2-\lambda_1}\left[\left|\psi(\vec p,-\lambda_1)\right\rangle\otimes\left|\psi(-\vec p,-\lambda_2)\right\rangle\right]\right\}
\end{eqnarray}
with
\begin{equation}
\Omega^J_{M,\lambda}[X]=\left[\frac{2J+1}{4\pi}\right]^{1/2}\int^{2\pi}_0d\phi\int^\pi_0d\theta\, \sin\theta\ {\cal D}^{J*}_{M,\lambda}(\phi,\theta,-\phi)\, R(\phi,\theta,-\phi)\, X(\phi,\theta),
\end{equation}
have the desired features, with $R(\alpha,\beta,\gamma)$ the rotation operator of Euler angles $\{\alpha,\beta,\gamma\}$ and ${\cal D}^{J}_{M,\lambda}(\alpha,\beta,\gamma)$ the Wigner $D$-matrices. Let us remark that $\left|\lambda_1,\lambda_2;J^P,M,\epsilon\right\rangle=\epsilon\, \left|-\lambda_1,-\lambda_2;J^P,M,\epsilon\right\rangle$; both states are equivalent up to an irrelevant $\epsilon$ factor.

The parity of the state~(\ref{hstate}) is given by $P=\epsilon\, \eta_1\eta_2(-1)^{J-s_1-s_2}$ \cite{jaco}, $\eta_i$ and $s_i$ being the intrinsic parity and spin of particle $i$ respectively. The parameter $\epsilon$, which can be equal to $\pm1$, fixes the final value of the parity. It is worth mentioning that, following the usual rules of spin coupling,
\begin{equation}
J\geq|\lambda_1-\lambda_2|.
\end{equation}
Another important point to notice is that when both particles have a spin degree of freedom, the helicity basis, spanned by the helicity states~(\ref{hstate}), is equivalent to an usual $\left|^{2S+1}L_J\right\rangle$ basis \cite{gie}. When at least one of the particles is massless, both basis are no longer equivalent but the helicity states can still be expressed as particular linear combinations of $\left|^{2S+1}L_J\right\rangle$ states \cite{jaco}. In order for our discussion to be complete, we have to mention that, in the case where the two constituent particles are identical, the helicity states have to be totally (anti)symmetrized, following the bosonic or fermionic nature of the particles. The action of the permutation operator on the helicity states can be found in Refs.~\cite{jaco,gluh1}, and we will not mention it here explicitly. But, it is worth saying that the symmetrization of the systems leads to constraints on the total angular momentum, thus to particular selection rules. This fact will be illustrated in the case of two-gluon glueballs.

A nice property of the quantum state~(\ref{hstate}) is that it is well defined for massless constituent particles. In particular, for a system made of two gluons, we have $\lambda_i=\pm1$ and $P=\epsilon(-1)^{J}$. An important quantity that can be computed in this case is the average square orbital angular momentum, that reads (see Ref.~\cite{gluh1} and Appendix~\ref{hsgl})
\begin{equation}\label{lsdef}
    \left\langle \vec L^{\, 2}\right\rangle=J(J+1)+2\lambda_1\lambda_2.
\end{equation}
A first idea to classify the glueballs is that $\left\langle \vec L^{\, 2}\right\rangle $ roughly sets the energy scale of a given state: The more rotational energy is contained in a glueball, the more the state is heavy. Notice that Eq.~(\ref{lsdef}) is valid for two-gluon states only.

The prescription to write a three-body helicity state has firstly been given in Ref.~\cite{wick3} and is based on an intuitive recoupling scheme. The first step is to write the two-body helicity state corresponding to, say, particles 2 and 3 in the center of mass frame of these two particles. This is done thanks to the result~(\ref{hstate}). Then, the Lorentz-boosted helicity state of the (2,3) cluster can be coupled to particle 1 with the same technique. The boost is needed because one wants the three-body helicity state to be expressed in the three-body rest frame, not in the rest frame of the (2,3) cluster. We propose the following notation to denote a three-body helicity state
\begin{equation}\label{h3g}
    \left| (\lambda_2,\lambda_3;j^\pi,m,\epsilon),\lambda_1;J^P,M,\rho\right\rangle.
\end{equation}
This last ket has to be understood as follows. Two constituent particles with helicities $\lambda_2$ and $\lambda_3$ are coupled to a given value of $j\geq|\lambda_2-\lambda_3|$, with a parity $\pi=\epsilon\eta_2\eta_3(-1)^{j-s_2-s_3}$ and a helicity $m$ following Eq.~(\ref{hstate}). Then the (2,3) cluster, seen as a pointlike massive particle with spin $j$ and helicity $m$, and the last gluon with helicity $\lambda_1$ are coupled to the total value of $J\geq|m-\lambda_1|$, with a parity $P=\rho\eta_1\pi(-1)^{J-j-s_1}=\epsilon\rho\eta_1\eta_2\eta_3(-1)^{J-s_1-s_2-s_3}$. This time, the product $\epsilon\rho$ determines the total parity, with $\rho=\pm1$. For a three-gluon system, the parity is given by $P=\epsilon\rho(-1)^J$. Although formally clear, the situation is much more complicated for three-body systems, because an infinity of helicity states of the form~(\ref{h3g}) can theoretically lead to a given $J^P$ state. Again, a classification can be made by assuming that a $J^{PC}$ three-gluon glueball is dominated by the helicity state for which the value of both the square orbital angular momentum of the cluster and the total square orbital angular momentum are minimal.

We propose to adopt a similar procedure to study four-gluon states: First, to write the helicity states for the (1,2) and (3,4) clusters and, second, to couple these two clusters. The advantage of doing so is that both clusters must be coupled as massive particles. Consequently, the usual $\left|^{2S+1}L_J\right\rangle$ basis can be used in this last step, and the effect of helicity is restricted to the two-gluon helicity states, which have been extensively studied in Ref.~\cite{gluh1}. Such a procedure can actually be extended to many-gluon or many-body systems, with an obvious increase of complexity with the number of constituent particles. Moreover, a global symmetry can become very difficult to implement when more than two constituent particles are identical (see for example Ref.~\cite{gross}, where the case of three identical fermions is considered).

Group theory generally gives powerful tools concerning the classification of hadrons and it will also be interesting to use it in the following. One could represent a constituent gluon by the Young diagram $\yng(1)^a$, the box denoting the fundamental representation of the little group $SO(3)$ \cite{tung} and $a=1,\dots,n$ being the color index associated with the gluon. Obviously $n=N^2_c-1$ with $N_c$ the number of colors. The isospin of a gluon is zero; this degree of freedom is thus trivial and will not be explicitly taken into account here. We then expect that the lightest glueballs made of $N_g$ constituent gluons will be those for which the $J^{PC}$ quantum numbers can be generated by the intrinsic color and spin degrees of freedom, i.e., since gluons are bosons, by the states appearing in the decompositions of the symmetrized tensor product
\begin{equation}\label{yt}
\left( \yng(1)^{a_1} \otimes \dots \otimes\ \yng(1)^{a_{N_g}} \right),
\end{equation}
where the symbols () denote the total symmetrization. Only the total spin $\vec J$ is a relevant observable, but the helicity states can always be decomposed in a usual $\left|^{2S+1}L_J\right\rangle$ basis. The $J^{PC}$ states appearing in the decomposition~(\ref{yt}) will consequently be those for which a $S$-wave component is present in the helicity state. Thus, we can expect that this will correspond to the lightest glueball states with $N_g$ gluons. Before illustrating these general ideas by considering explicit examples, we refer the reader to Ref.~\cite{fulh} for a discussion about representation theory. More precisely, the symmetrized tensor products -- demanded by the Pauli principle -- that arise when doing tensor products of several identical boxes $\yng(1)^a$ are referred to as plethysms. It is worth mentioning Ref.~\cite{glusp}, in which a fully group-theoretical classification of glueballs is performed following a $SU(3)\times SU(2)$ scheme. But, the influence of gluon helicity through the helicity formalism has, to our knowledge, never been taken into account and leads to interesting and new results as we will see through this paper.

\section{The glueball spectrum}\label{sec2}

\subsection{Two-gluon states}

Bound states of two gluons are the most studied purely gluonic systems in the literature. Since their properties are rather well-known, it is useful to re-analyze them using the method introduced in Sec.~\ref{gene} in order to check its compatibility with previous works. Let us begin by a Young diagrams analysis. For a system made of two constituent gluons, one has
\begin{equation}\label{y1}
\left( \yng(1)^{a} \otimes \yng(1)^{b} \right) = \left\{ \yng(2) \oplus \bullet\right\}^{(ab)}
\oplus \yng(1)^{[ab]},
\end{equation}
where $(a_1 a_2 \ldots a_{N_g})$ and $[a_1 a_2 \ldots a_{N_g}]$ denote respectively the totally symmetric and antisymmetric color wave functions. The notation {\{\}} allows avoiding to repeat the same color superscript for different spin irreps, and the $\bullet$ stands for a spin zero state. Let us briefly make a comment about the dimension of the symbols appearing in this last equality. The Young tableau $\yng(1)^a$ has a dimension $3n$, i.e. the product of the spin and color degrees of freedom. Then, the left hand side of Eq.~(\ref{y1}) has the dimension $3n(3n+1)/2$ since only its totally symmetric part is considered. The three terms appearing on the right hand side have dimension $5n(n+1)/2$, $n(n+1)/2$, and $3n(n-1)/2$ respectively, the sum of which gives $3n(3n+1)/2$ as expected.

Since a two-gluon state must be in a color singlet, its color wave function is trivially $[\bm 8,\bm 8]^{{\bm 1}_s}$, the resulting singlet being symmetric under the permutation of both gluons. Only the first two terms of the decomposition~(\ref{y1}) then correspond to colorless glueballs: An antisymmetric color wave function cannot correspond to a color singlet. Notice that the $\yng(1)^{[ab]}$ state could correspond to colored glueballs with the color wave function $[\bm 8,\bm 8]^{{\bm 8}_a}$, $[\bm 8,\bm 8]^{{\bm 10}_a}$ or $[\bm 8,\bm 8]^{{\overline{{\bm 10}}}_a}$, and could become relevant in a deconfined medium as the quark-gluon plasma. Notice that the superscript $\bm X_{a/s}$ is used to denote a symmetric ($s$) or antisymmetric ($a$) representation $\bm X$. The first two states of the decomposition~(\ref{y1}) are $J^C=2^{+}$ and $0^{+}$ glueballs respectively. The $C$-parity of a two-gluon system is always positive as shown in Appendix~\ref{cval}. We do not discuss the parity of these state at this stage since only the helicity formalism can bring relevant information about it.

The helicity states corresponding to two-massless gluon systems can be explicitly written thanks to Eq.~(\ref{hstate}). Since all the necessary calculations have been performed in Ref.~\cite{gluh1}, we only mention here the final results concerning the two-gluon helicity states and their properties. We have
\begin{eqnarray}\label{2ghs}
\left|1,1;(2k)^+,M,1\right\rangle&\Rightarrow& 0^{++},\, 2^{++},\, 4^{++},\dots\nonumber\\
\left|1,1;(2k)^-,M,-1\right\rangle&\Rightarrow& 0^{-+},\, 2^{-+},\, 4^{-+},\dots\\
\left|1,-1;(2k+2)^+,M,1\right\rangle&\Rightarrow& 2^{++},\, 4^{++},\, 6^{++},\dots\nonumber\\
\left|1,-1;(2k+3)^+,M,-1\right\rangle&\Rightarrow& 3^{++},\, 5^{++},\, 7^{++},\dots\nonumber,
\end{eqnarray}
with $k=0$, 1, \dots The selection rules that appear on the total spin $J$ are actually a consequence of the fact that these helicity states have to be totally symmetric \cite{gluh1}. In particular, the state $\left|1,-1;2^-,M,-1\right\rangle$ is forbidden for such a reason. In agreement with Eq.~(\ref{y1}) and Fig.~\ref{fig1}, the lightest glueballs correspond to $\left|1,1;0^+,0,1\right\rangle$ and $\left|1,-1;2^+,M,1\right\rangle$ respectively, for which the square orbital angular momentum is $2$ and $4$ respectively. The $0^{-+}$ glueball, given by $\left|1,1;0^-,0,-1\right\rangle$, is heavier than the $0^{++}$ one although its square orbital momentum is also equal to $2$. This shows that other effects come into play into glueballs: We suggested in Ref.~\cite{gluh1} that the mass splitting between the scalar and pseudoscalar glueball is due to instanton-induced interactions. Finally, always by looking at Eq.~(\ref{lsdef}), we are led to the conclusion that the $2^{-+}$ is lighter than the $3^{++}$ one, as observed in lattice QCD (see Fig.~\ref{fig1}).

An important remark is that no $J^{PC}=1^{\pm+}$ glueball should be present in the energy range of the currently observed $C=+$ states because of the selection rules on the states~(\ref{2ghs}). This is indeed the case in lattice QCD and gives support to the framework that we propose. We notice that this result is obviously similar to Yang's theorem stating that a vector meson cannot decay into two photons \cite{yang}. But, if constituent gluons are spin-1 particles, low-lying $1^{\pm+}$ states are then allowed, in disagreement with what is found in lattice QCD \cite{gluh1} -- consider the $\left|^3P_1\right\rangle$ state for example. The presence of such vector glueballs is actually due to the fact that $\lambda_i=0$ is allowed with spin-1 constituent gluons.

\subsection{Three-gluon states}

When three constituent gluons are taken into account, the following decomposition holds
\begin{equation}\label{y3}
\left( \yng(1)^{a} \otimes \yng(1)^b \otimes \yng(1)^{c} \right) = \left\{ \yng(3) \oplus \yng(1) \right\}^{(abc)}
\oplus \left\{ \yng(2) \oplus \yng(1) \right\}^{M^{|2,1|}_{abc}} \oplus \bullet^{[abc]},
\end{equation}
where $M^{|2,1|}_{abc}$ denotes a color configuration with the mixed symmetry $\yng(2,1)$. But, it can be checked that no color singlet can be made with this mixed symmetry (see Appendix~\ref{cval}). So, the third and fourth states of the decomposition~(\ref{y3}) cannot be colorless three-gluon states. But, the symmetrical color wave function $(abc)$ corresponds to the following color singlet, $[\left[\bm 8,\bm 8\right]^{\bm 8_s},\bm 8]^{\bm 1_s}$, and leads to $C=-$ glueballs. Similarly, the antisymmetrical color wave function $[abc]$ corresponds to the following color singlet, $[\left[\bm 8,\bm 8\right]^{\bm 8_a},\bm 8]^{\bm 1_a}$, and leads to $C=+$ glueballs as already pointed out in Ref.~\cite{mink}. Consequently, the glueballs corresponding to the first two states of decomposition~(\ref{y3}) are $J^C=3^{-}$ and $1^{-}$ states respectively, while the last state is a $0^{+}$ one. We will actually focus on the $C=-$ three-gluon states in this section since no mixing of these states with two-gluon ones can exist. We remark that $C=-$ three-gluon glueballs are formally identical to three-photon states.

Following the arguments that we gave in the previous section, the lightest three-gluon states should be those for which the two square orbital momenta appearing in the three-body system have both a minimal value. Ignoring possible instanton effects, the most obvious possibility to build such a system is the following
\begin{equation}\label{3ghs}
    \hat{\cal S} \left| (1,1;0^\epsilon,0,\epsilon),1;1^{-\epsilon\rho},M,\rho\right\rangle,
\end{equation}
where $j$ and $J$ have the minimal allowed value, leading to $1^{\pm-}$ glueballs, and where $\hat{\cal S}$ is the three-body symmetrizer. We stress that, formally, an infinity of $\{j,m,\epsilon,\rho\}$ sets can lead to a given $J^{PC}$ state. For example, the quantum state of the $1^{+-}$ glueball is rigorously given by a linear combination of all the totally symmetric three-body helicity states for which the values of $\{j,m,\epsilon,\rho\}$ can lead to $J^{PC}=1^{+-}$ state. Here we assume that the lowest-lying glueballs are dominated by the component in which the minimal amount of rotational energy is present. Similarly, light $J^{PC}=3^{\pm -}$ glueballs can a priori be built from the following state,
\begin{equation}\label{3ghs2}
    \hat{\cal S} \left| (1,-1;2^+,-2,1),1;3^{-\rho},M,\rho\right\rangle,
\end{equation}
where again both $j$ and $J$ are minimal following the values of $\lambda_i$ and $m$ that have been chosen, and where the rotational energy is the lowest possible for the cluster (2,3).

In order to understand the mass hierarchy of the lattice QCD states, we propose to introduce the notion of natural and unnatural parity for glueballs. It is indeed known in hadronic and nuclear physics that states with natural parity are lighter than those with unnatural parity. We suggest that the natural parity of a glueball made of $N_g$ gluons is given by $(-1)^{N_g}\, (-1)^J$, where the first factor comes from the intrinsic parity of the $N_g$ gluons, and where the second one is the expected parity of a state with spin $J$. For two-gluon glueballs, the $0^{++}$ and $2^{++}$ states have a natural parity, and are lighter than the $0^{-+}$ and $2^{-+}$ ones. In the case of three-gluon systems, the introduction of natural parity leads to the prediction that the $1^{+-}$ ($3^{+-}$) glueball should be lighter than the $1^{--}$ ($3^{--}$) one. Thus, we predict the following inequalities concerning the glueball masses
\begin{equation}
M_{1^{+-}}<M_{1^{--}},\quad M_{3^{+-}}<M_{3^{--}}.
\end{equation}
This is indeed what is found in lattice QCD, as one can see in Fig.~\ref{fig1}. We mention also the inequality $M_{1^{\pm -}}<M_{3^{\pm -}}$, which can be deduced from the angular momentum content of the states~(\ref{3ghs}) and (\ref{3ghs2}).

An important point to notice is that no low-lying $0^{\pm -}$ glueball can be obtained since the lowest allowed value for $J$ is 1. This result has already been mentioned in Ref.~\cite{fumi}, where it is shown that no (pseudo)scalar three-photon state can exist, excepted in very asymmetric configurations that seem irrelevant to describe bound states. The $0^{+-}$ glueball that is observed in lattice QCD should then be seen as either a highly excited three-gluon state or as a low-lying state made of at least four-gluons (two gluons are forbidden by the negative $C$-parity). In both cases its large mass is explained (it is the heaviest glueball that is currently known in lattice QCD). But, as we will see in the following sections, the hypothesis of a low-lying four-gluon state appears to be particularly relevant. We also remark that, when spin-1 constituent gluons are assumed, the trivial state $\left| (0,0;0^\epsilon,0,\epsilon),0;0^{-},0,\epsilon\right\rangle,$ can lead to a $0^{--}$ glueball which should be among the lightest ones in the $PC=--$ channel, but which is not seen in lattice QCD. This favors the helicity-1 approach since the $C=-$ sector is known in a quite large mass range, and no $0^{--}$ glueball has been observed yet. It is finally worth mentioning that with spin-1 gluons, a $0^{+-}$ three-gluon glueball is also allowed, but it has been shown in a previous work that its mass is far too close of the other three-gluon candidates with respect to the mass splittings that are observed in lattice QCD \cite{gvm2}.

The additional $2^{--}$ and $2^{+-}$ states that are present in Fig.~\ref{fig1} could be understood as the first excitations of the helicity state~(\ref{3ghs}) for example. By excited state, we mean a state whose total spin $J$ is higher than the lowest allowed value. Since the parity is proportional to $(-1)^J$, the $2^{--}$ ($2^{+-}$) glueball could be an excitation of the $1^{+-}$ ($1^{--}$) glueball. This point is still compatible with lattice QCD since the $2^{--}$ ($2^{+-}$) glueball is observed to be heavier than the $1^{+-}$ ($1^{--}$) glueball. Notice that $M_{2^{--}}<M_{2^{+-}}$ is observed: The glueball with natural parity is still the lightest. It could be argued that states such as $\left|(1,-1;2^+,1,1),-1;2^{\rho},M,\rho\right\rangle $ could lead to $2^{\pm -}$ glueballs without any excitation. But the problem actually comes from the value $m=1$ of the cluster helicity. Indeed, if $m=\pm1$, low-lying glueballs such as $\left|(1,-1;2^+,1,1),1;0^{\rho},0,\rho\right\rangle$ are allowed. This would be in contradiction with the results of Ref.~\cite{fumi}, forbidding (pseudo)scalar totally symmetric three-gluon systems, or at least low-lying ones. We actually think that the case $|m|=1$ is forbidden by the requirement that the three-body helicity states have to be totally symmetrized. A confirmation of this point needs an explicit implementation of the three-body symmetrizer and a computation of its action on the helicity states. Such a task is far from being trivial (see Ref.~\cite{gross} for a fermionic example), and is out of the scope of the present paper. Nevertheless, we suggest that $2^{\pm-}$ three-gluon glueballs are most likely excited states.

\subsection{Four-gluon states}\label{4ganaly}

The situation becomes a little bit more involved in the case of four-gluon systems. One obtains
\begin{eqnarray}\label{y4}
\left( \yng(1)^{a} \otimes \yng(1)^{b} \otimes \yng(1)^{c} \otimes \yng(1)^{d} \right) &=&
\left\{ \yng(4) \oplus \yng(2) \oplus \bullet \right\}^{(abcd)} \oplus
\left\{ \yng(2) \oplus \bullet \right\}^{M^{|2,2|}_{abcd}} \nonumber \\
&& \oplus \yng(1)^{M^{|2,1,1|}_{abcd}} \oplus
\left\{ \yng(3) \oplus \yng(2) \oplus \yng(1) \right\}^{M^{|3,1|}_{abcd}},
\end{eqnarray}
where $M^{|3,1|}_{abcd}$, $M^{|2,2|}_{abcd}$ and $M^{|2,1,1|}_{abcd}$ denote the color configurations with the symmetries
$\yng(3,1)$, $\yng(2,2)$, and $\yng(2,1,1)$ respectively. It is worth discussing a bit the different allowed color singlets in the case of four-gluon systems. By using a (1,2)$-$(3,4) recoupling scheme as it is done within the helicity formalism, we can write all the confined color singlets as follows: $[[\bm 8,\bm 8]^{\bm 8_a},[\bm 8,\bm 8]^{\bm 8_a}]^{\bm 1}$, $[[\bm 8,\bm 8]^{\bm 8_a},[\bm 8,\bm 8]^{\bm 8_s}]^{\bm 1}$, $[[\bm 8,\bm 8]^{\bm 8_s},[\bm 8,\bm 8]^{\bm 8_s}]^{\bm 1}$, $[[\bm 8,\bm 8]^{\overline{\bm{10}}_a},[\bm 8,\bm 8]^{\bm{ 10}_a}]^{\bm{1}}$, and $[[\bm 8,\bm 8]^{\bm{27}_s},[\bm 8,\bm 8]^{\bm{ 27}_s}]^{\bm{1}}$ (the $[[\bm 8,\bm 8]^{\bm 1},[\bm 8,\bm 8]^{\bm 1}]^{\bm 1}$ color wave function would lead to a deconfined four-gluon state which would obviously split into two two-gluon glueballs). These color singlets lead to different values of the $C$-parity that have been computed in Appendix~\ref{cval}. We remark first that no colorless glueball can be made with the $M^{|3,1|}$ symmetry. Only the first six irreps of the decomposition~(\ref{y4}) are thus relevant to our purpose. The corresponding states can be identified with $J^C=\{4,\, 2,\, 0\}^+$, $\{2,\, 0\}^{\pm}$, and $1^-$ glueballs respectively. Due to the many configurations that are possible, an exhaustive analysis of four-gluon states would be rather tedious, and not very worthy since only the $0^{+-}$ glueball appears as a relevant four-gluon candidate in lattice QCD. In this section, we rather show that a four-gluon $0^{+-}$ glueball is allowed and we focus on its specific properties.

It is known that, at tree level, the short-range interactions between two constituent gluons are mainly proportional to $C_{gg}-6$, where $C_{gg}$ is the quadratic Casimir of $\mathfrak{su}(3)$ in the representation of the gluon pair \cite{oge}. The configurations in which a gluon pair is in the $\bm{10}_a$ or $\overline{\bm{10}}_a$ representation would lead to vanishing short-range interactions of one-gluon-exchange type, while the $\bm{27}_s$ representation would lead to repulsive short-range forces. Color wave functions in which the gluon pairs are in a color octet seem thus favored since the short-range interactions are attractive in this case. Configurations involving the $\bm{27}_s$ representation should conversely lead to heavier glueballs because of the existence of repulsive forces. It can be checked that, among the color wave functions involving color octets only, $[[\bm 8,\bm 8]^{\bm 8_a},[\bm 8,\bm 8]^{\bm 8_s}]^{\bm 1}$ is the unique configuration that has a negative $C$-parity. But, the resulting color singlet can only have the mixed permutation symmetry $M^{|2,1,1|}$, that cannot be associated with a low-lying $J=0$ glueball as suggested by Eq.~(\ref{y4}). A light $0^{+-}$ glueball can consequently not be built from color configurations where each gluon pair is in a color octet; higher representations are needed. As it is shown in Appendix~\ref{cval}, a color wave function of the type $[[\bm 8,\bm 8]^{\overline{\bm{10}}_a},[\bm 8,\bm 8]^{\bm{ 10}_a}]^{\bm{1}}$ can lead to a $C=-$ glueball with the $M^{|2,2|}$ permutation symmetry. Consequently, the unique $0^{\pm -}$ state that can be built from Eq.~(\ref{y4}) corresponds to $\bullet^{M^{|2,2|}_{abcd}}$. Let us explicitly build a $0^{+-}$ state by using the helicity formalism.

One has first to remember that the $0^{+-}$ four-gluon glueball should have a particular mixed helicity symmetry of the form $\yng(2,2)$. Let us briefly come back to the color wave function. It reads $[[\bm 8_1,\bm 8_2]^{\bm{10}_a},[\bm 8_3,\bm 8_4]^{\overline{\bm{10}}_a}]$, where indices labeling the constituent gluons have been added. In this recoupling scheme, the helicity states must have precisely the mixed symmetry $\scriptsize{\young(13,24)}$ in order for the Pauli principle to be satisfied. Both gluon clusters are thus totally antisymmetric, a configuration that is forbidden for two-gluon glueballs. Let us define the antisymmetrizer of the cluster $(i,j)$ by $\hat{\cal A}=(1-P_{ij})/2$, $P_{ij}$ being the permutation operator. Then it can be deduced from the results of Ref.~\cite{gluh1} that the action of the antisymmetrizer on the two-gluon helicity states is given by
\begin{subequations}
\begin{eqnarray}
2\hat{\cal A}\left|1,1;J^P,M,\epsilon\right\rangle&=&\left[1-(-1)^J\right]\left|1,1;J^P,M,\epsilon\right\rangle,\\
2\hat{\cal A}\left|1,-1;J^P,M,\epsilon\right\rangle&=&\left[1-\epsilon(-1)^J\right]\left|1,-1;J^P,M,\epsilon\right\rangle,
\end{eqnarray}
\end{subequations}
leading to selection rules constraining the total angular momentum. Remember that we are looking for the lightest $0^{+-}$ state: Both gluons clusters should be as light as possible. By looking at Eq.~(\ref{lsdef}), one can observe that the lightest antisymmetric two-gluon clusters correspond to the states $\left|1,1;1^{-\epsilon},M,\epsilon\right\rangle$ and $\left|1,-1;2^{-},M,-1\right\rangle$, for which $\left\langle \vec L^{\, 2}\right\rangle=4$. Such a vector state is coherent with Eq.~(\ref{y1}).

Now that the quantum state of the clusters is known, we can lastly couple them in order to obtain a $0^{+-}$ state (the negative $C$-parity is guaranteed by the color wave function). In order to satisfy at best the mixed symmetry of the four-gluon state, we will ask for the two clusters to be identical and for the two-cluster state to be the lightest $J=0$ one which is totally symmetric. This state is simply the $\left|^1S_0\right\rangle$ one, either spin-1 or spin-2 clusters are assumed. Notice that the parity of the two-cluster state is $(\pm 1)^2\, (-1)^{L}=+1$ as demanded. This is also the natural parity of a $J=0$ four-gluon glueball.

\section{A link with gluelumps}\label{sec3}

As it was recalled in the introduction, gluelumps are bound states of the gluonic field in a static color octet source. This color octet source is assumed to have the quantum numbers of the vacuum, i.e. $J^{PC}=0^{++}$. Following the calculations of Ref.~\cite{gl1}, the lowest-lying gluelumps are lighter than all the currently known glueballs in lattice QCD. In consequence, we propose to identify the lowest-lying gluelumps with states made of only one constituent gluon (the static source holds for a scalar particle in a color octet, fixed at the center of mass). Such one-gluon states are generated by the tensor product
\begin{equation}\label{glu1}
\yng(1)^a \otimes \bullet^b = \yng(1)^{(ab)} \oplus \yng(1)^{[ab]},
\end{equation}
where $\bullet^b$ represents the color octet static source. Note that no symmetrization must be achieved between a gluon and a static color source. As seen in Appendix~\ref{cval}, only the symmetric color state $(ab)$ can be reached by a color singlet, with $C=-$ this time. The corresponding helicity states are given by $\left|1,0;J^P,M,\epsilon\right\rangle$, with a parity given by $P=\epsilon(-1)^J$ as the static source is assumed to have a positive intrinsic parity.  We notice that $J\geq 1$, so that the two lightest gluelumps should be the $1^{+-}$ and $1^{--}$ ones, followed by their first excitations, i.e. the $2^{--}$ and $2^{+-}$ states respectively. If the lattice QCD gluelump states are ordered by increasing mass, one finds the following quantum numbers: $1^{+-},\, 1^{--},\, 2^{--},\, 2^{+-},\, 3^{+-},\, 0^{++},\dots$ \cite{gl2}. This ordering is coherent with the one-gluon picture and with the natural parity criterion until a $0^{++}$ state is reached, because such a positive $C$-parity state cannot be reached with a single constituent gluon.

As pointed out in Ref.~\cite{sczg}, a $0^{++}$ gluelump should at least contain two constituent gluons. Let us check this point within our formalism. The lowest-lying two-gluon gluelumps are expected to be generated by the following tensor product
\begin{equation}\label{glus2}
\left( \yng(1)^{a} \otimes \yng(1)^{b} \right) \otimes \bullet^c = \left\{ \yng(1) \oplus \bullet \right\}^{(abc)}
\oplus \left\{ \yng(2) \oplus \yng(1) \oplus \bullet \right\}^{M^{|2,1|}_{abc}} \oplus \yng(1)^{[abc]},
\end{equation}
where the $\bullet^{(abc)}$ state with the natural parity should be the lowest-lying two-gluon gluelump. It indeed corresponds to a $0^{++}$ state of the form
\begin{equation}
    \hat{\cal S}_{23}\left|(1,1;0^\epsilon,0,\epsilon);0;0^+,0,\epsilon\right\rangle,
\end{equation}
where the two gluons have been coupled first and symmetrized. Such a state has the lowest possible values for both the internal and total angular momenta and is similar to a two-gluon glueball. In particular, its mass is equal to $r_0M_{0^{++}}=5.02\pm0.46$ \cite{gl1}, that is a mass which lies in the typical range of the low-lying two-gluon glueballs (see Fig.~\ref{fig1}).

When gluons have a spin-1, a light $0^{--}$ one-gluon gluelump can be simply built from $\left|0,0;0^-,0,-1\right\rangle$, that is a single gluon with the zero spin projection. Interestingly, no $0^{--}$ gluelump has been observed yet in lattice QCD, still favoring the helicity-1 picture that we develop in the present paper.

\begin{table}[b]
\caption{Left: Natural parity low-lying $N_g$-gluon candidates. The quantum numbers and the masses of these states are denoted by $J^{PC}$ and $r_0M_{J^{PC}}$ respectively. Averaged masses are computed following Eq.~(\ref{mav}). The $N_g=1$ states are taken from Ref.~\cite{gl1} while the other data are taken from Ref.~\cite{lat3}. Masses are given in lattice units. Right: \textit{Idem} for the unnatural parity candidates.}
\setlength{\extrarowheight}{4pt} \begin{tabular}{ccccccccc}
  \hline\hline
 $N_g$ && $J^{PC}$ & $r_0M_{J^{PC}}$ \cite{lat3,gl1} & $r_0\overline{ M}(N_g,1)$ && $J^{PC}$ & $r_0M_{J^{PC}}$ \cite{lat3,gl1} & $r_0\overline{ M}(N_g,-1)$ \\
\hline
1 && $1^{+-}$ & 2.25$\pm$0.39 & 2.25$\pm$0.39 && $1^{--}$ & 3.18$\pm$0.41 & 3.18$\pm$0.41\\
2 && $0^{++}$ & 4.16$\pm$0.15 & 5.55$\pm$0.12 && $0^{-+}$ & 6.25$\pm$0.12& 7.23$\pm$0.14\\
  && $2^{++}$ & 5.83$\pm$0.11 & && $2^{-+}$ & 7.42$\pm$0.14 &  \\
3 && $1^{+-}$ & 7.27$\pm$0.11 & 8.33$\pm$0.12 &  &$1^{--}$ & 9.34$\pm$0.13& 9.98$\pm$0.14\\
  && $3^{+-}$ & 8.79$\pm$0.12 & &&$3^{--}$ & 10.25$\pm$0.14\\
4 && $0^{+-}$ & 11.66$\pm$0.19 & 11.66$\pm$0.19  &&  & \\
\hline\hline
\end{tabular}
\label{tab1}
\end{table}

\section{Mass hierarchy of glueballs and gluelumps}\label{sec4}

\begin{figure}[t]
  \includegraphics[width=8.5cm]{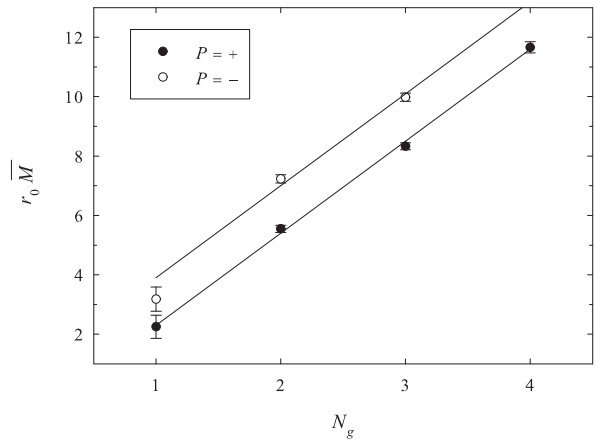}
\caption{Plot of the spin-averaged masses of the natural parity (full circles) and unnatural parity (empty circles) states of Table~\ref{tab1} versus the presumed number of constituent gluons. Equation~(\ref{mav2}) is plotted with the optimal values~(\ref{param}) for $\beta$ and $\gamma$. }
\label{fig2}
\end{figure}

We have observed through the previous sections that the different lowest-lying one-, two-, three- and four-gluon states can be separated into glueballs with natural parity or unnatural parity. The corresponding low-lying glueballs that are currently observed in lattice QCD are listed in Table.~\ref{tab1}. The left part is concerned with the natural parity glueballs ($P=+$) while the right part is concerned with the unnatural parity states ($P=-$). In order to explain the mass hierarchy of these states, we define the spin-averaged mass as usual by
\begin{equation}\label{mav}
\overline{M}(N_g,P)=\frac{\sum_J (2J+1) M_{J^{PC}} }{\sum_J (2J+1)},
\end{equation}
where $M_{J^{PC}}$ is the mass of the glueball which is a $N_g$-gluon candidate with $J^{PC}$ quantum numbers. The case $N_g=1$ is the gluelump case, while $N_g>1$ corresponds to glueballs. It appears that the spin-averaged masses grow linearly with the presumed number of constituent gluons; we then propose to fit the data of Table~\ref{tab1} with the following form
\begin{equation}\label{mav2}
r_0\overline{M}(N_g,P)=\beta\, N_g-\gamma\, P.
\end{equation}
A linear regression on the natural parity states leads to
\begin{equation}\label{param}
\beta=3.10\pm0.08,\quad {\rm and}\quad \gamma=0.81\pm0.21.
\end{equation}
As it can be observed in Fig.~\ref{fig2}, formula~(\ref{mav2}) with the fitted values~(\ref{param}) reproduces rather well the spin-averaged masses. It is worth mentioning that this last mass formula predicts a $0^{--}$ four-gluon glueball in the mass range $r_0\, M_{0^{--}}=13.21\pm0.53$.

Formula~(\ref{mav2}) actually gives support to the constituent gluon picture, each constituent gluon roughly bringing to the total glueball mass an amount which is equal to $\beta$. Additional mechanisms (instanton-induced interactions, spin-dependent forces, etc.) are expected to come into play as higher-order terms and are all contained in the $-\gamma\, P$ term. If the particular value of $\gamma$ probably needs a very accurate model to be understood, $\beta$ can be obtained within the framework of the flux tube model. Indeed, $\beta$ can be simply obtained from formula~(\ref{mav2})
\begin{equation}\label{bgdef}
    \beta=\frac{1}{2 N_g}\left[r_0\overline{M}(N_g,1)+r_0\overline{M}(N_g,-1)\right].
\end{equation}
The most simple case is the gluelump one, for which $N_g=1$. $\beta$ is then related to a typical gluelump mass, that can be obtained from a gluelump Hamiltonian in the flux tube model. At the dominant order, it reads
\begin{equation}\label{gluelu}
    H=\sqrt{\vec p^{\, 2}}+ar-3\frac{\alpha_s}{r},
\end{equation}
where $\vec p$ and $\vec r$ are conjugate variables and where $r=|\vec r\,|$ is the separation between the constituent gluon and the static source.
It is spin-independent but, in our formalism, the effect of the gluons helicity is included at the level of the wave functions, not in the Hamiltonian itself. Although we do not use an explicitly covariant formalism, the semirelativistic kinetic term $\sqrt{\vec p\,^2}$ should be relevant to describe the dynamics of the constituent gluon.
$a$ is the energy density of the flux tube linking this gluon to the static source, and $\alpha_s$ is the strong coupling constant. The flux tube actually comes from the nonperturbative part of the QCD interactions and generates the confinement, while the Coulomb part is the dominant part of the one-gluon-exchange potential between the constituent gluon and the static source \cite{oge}. Assuming the Casimir scaling hypothesis, one can write $a=(9/4)\sigma$ with $\sigma$ the energy density of the flux tube in a meson. In order to compare our results with lattice QCD, we can define $\sigma=1/r^2_0$, and work with the new coordinates $\vec q=r_0\, \vec p$ and $\vec x=\vec r/r_0$. Then, Hamiltonian~(\ref{gluelu}) becomes
\begin{equation}\label{gl2}
    r_0\, H=\sqrt{\vec q\,^2}+\frac{9}{4}x-3\frac{\alpha_s}{x}.
\end{equation}
Following Eq.~(\ref{bgdef}) and Table~\ref{tab1}, $\beta$ should be given by $(M_{1^{+-}}+M_{1^{--}})/2$, both masses being computed from this last Hamiltonian. As mentioned in Sec.~\ref{sec3}, two helicity states correspond to a single gluon with helicity-1: $\left|\Psi_\pm\right\rangle=\left|1,0;J^P,M,\pm\right\rangle$, with $J\geq1$ (see Appendix~\ref{hsgl}). The eigenstates of Hamiltonian $H$ can be accurately computed using the Lagrange-mesh method \cite{sem01}, using $\left\langle\Psi_\pm\right| \vec L^{\, 2}\left|\Psi_\pm\right\rangle=J(J+1)$. The ground state of Hamiltonian~(\ref{gl2}) is then obtained for $J=1$ and $n=0$ ($n$ is the radial quantum number), and it is clear that $M_{1^{+-}}=M_{1^{--}}$ with this simple model. Let us denote $M_{nJ}$ the eigenvalues of Hamiltonian~(\ref{gl2}). One gets, after a numerical evaluation, $M_{01}=3.09\approx\beta$ for $\alpha_s=0.32$. This value of the strong coupling constant is quite usual in potential models. This achieves to clarify the physical content of the coefficient $\beta$. An important point to stress is that, in a usual, spin-based model, the lowest allowed value for $\left\langle \vec L^{\, 2}\right\rangle$ is 0 and would lead to the conclusion that $\beta$ has to be identified with $M_{00}$. But, it is readily checked that $M_{00}\rightarrow-\infty$ unless $\alpha_s=0$, which is not physically relevant. In this case, the use of spin-1 gluons clearly leads to unphysical behaviors.

An estimation of $\beta$ can also be obtained by considering the case $N_g=2$. Particularized to two-gluon glueballs, the flux tube Hamiltonian~(\ref{gl2}) simply becomes
\begin{equation}\label{gl3}
    r_0\, H=2\sqrt{\vec q\,^2}+\frac{9}{4}x-3\frac{\alpha_s}{x},
\end{equation}
were the kinetic parts of two gluons have been included. In the same way than for gluelumps, $\beta$ can be computed from the eigenstates of Hamiltonian~(\ref{gl3}). In particular, the knowledge of the $0^{\pm+}$ and $2^{\pm+}$ glueball masses finally lead us to a well-defined value for $\beta$. We refer the reader to Ref.~\cite{gluh1} for a detailed discussion about the computation of the two-gluon glueball spectrum within the helicity formalism. Once all calculations are done, with $\alpha_s=0.32$ as for gluelumps, we find $\beta=3.33$, which is not far from the optimal value.

\section{The $0^{+-}$ four-gluon glueball}\label{4glu}

We have outlined in Sec.~\ref{4ganaly} the fact that the $0^{+-}$ glueball, which is currently the heaviest one that has been observed in lattice QCD, is a relevant four-gluon glueball candidate. We have also shown that such a system has a particular structure: Its color wave function should be $[[\bm 8,\bm 8]^{\bm{10}_a},[\bm 8,\bm 8]^{\overline{\bm{10}}_a}]^{\bm{1}_-}$ (see Appendix~\ref{cval}), with a mixed permutation symmetry of the form $\yng(2,2)$. To our knowledge, it is the first time that a four-gluon interpretation of a lattice QCD glueball is proposed. A confirmation of this point is then important, in particular by performing a quite crude but physically relevant four-body calculation of the $0^{+-}$ mass within a constituent gluon approach and the helicity formalism.

\subsection{Hamiltonian}

As in Hamiltonians used above, we adopt the flux tube picture stating that straight flux tubes start from the constituent particles and meet in such a way that the energy of the system is minimal. It was suggested in the recent lattice QCD studies of Refs.~\cite{model} that the less energetic configuration for many-gluon systems (more than two) is the one in which each constituent gluon is linked to its two nearest neighbors by two fundamental flux tubes. Phenomenological arguments also support this picture \cite{math}. In agreement with this idea, we suggest that the Hamiltonian describing a glueball made of four gluons is given at the dominant order by
\begin{equation}\label{mainH0}
    H=\sum^4_{i=1}\sqrt{\vec p\,^2_i}+\sigma\sum_{{\rm Cycle}}|\vec r_i-\vec r_{i+1}|,
\end{equation}
where $\vec r_i$ and $\vec p_i$ are the position and momentum of particle $i$ respectively. The only parameter of this Hamiltonian is the energy density of a fundamental flux tube. Using the dimensionless variables introduced in Sec.~\ref{sec4}, Eq.~(\ref{mainH0}) can be recast in the form
\begin{equation}\label{mainH}
r_0\, H=\sum^4_{i=1}\sqrt{\vec q\,^2_i}+\sum_{{\rm Cycle}}|\vec x_i-\vec x_{i+1}|,
\end{equation}
in which no free parameter is present.

Hamiltonian~(\ref{mainH}) seems too simple at first sight. Indeed, pair interactions coming from one-gluon-exchange processes have not been included. However, it has be shown in Ref.~\cite{oge} that the relativistic corrections to bring to the confining potential are proportional to two color factors. One of them is a projector on a color octet state and vanishes in our case because a gluon pair is always in a color (anti)decuplet. The other color factor also vanishes because it is proportional to $C_{gg}-6$, which vanishes in the color (anti)decuplet channel. Remarkably, the few physical ingredients that have been put in Hamiltonian~(\ref{mainH}) appear to be enough to accurately describe the $0^{+-}$ glueball thanks to the vanishing of the relativistic corrections in this case. It is possible that other nonperturbative effects like instanton-induced interactions could contribute in a non trivial way, but we neglect them in this simple first approach.

\begin{figure}[t]
  \includegraphics[width=5.5cm]{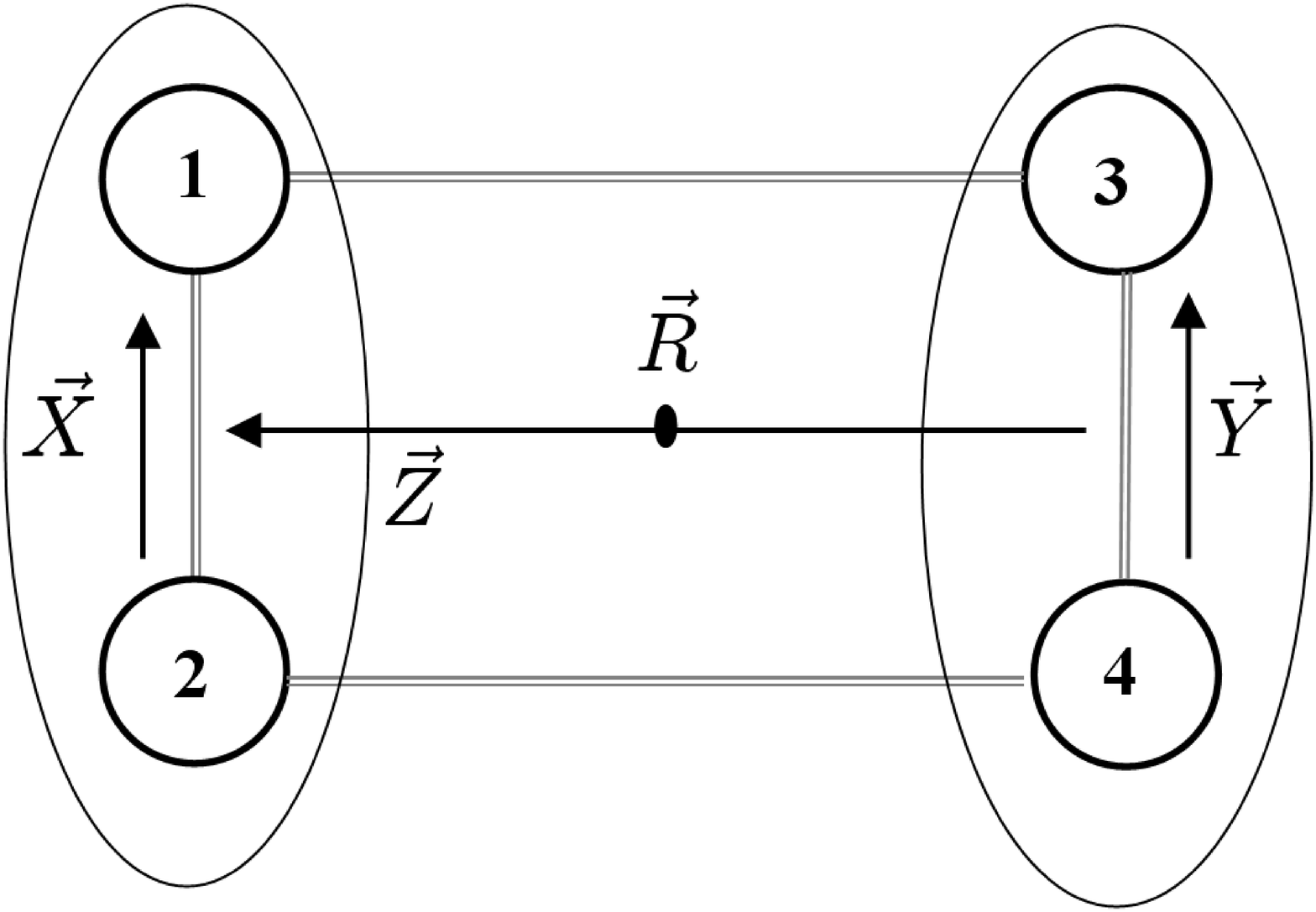}
\caption{Schematic representation of the four-gluon system considered. Four fundamental flux tubes (gray lines) link the constituent gluons (circles), illustrating Hamiltonian~(\ref{mainH}). The relative coordinates defined by relations~(\ref{relco}) are also represented (black arrows), while the center of mass, denoted as $\vec R$, is given by Eq.~(\ref{cm}). The (1,2) and (3,4) gluon clusters have been surrounded.}
\label{fig3}
\end{figure}

For future computations, it will be very convenient to have an appropriate set of relative coordinates at our disposal. We first define $\vec R$ the center of mass of the system as
\begin{equation}\label{cm}
    \vec R= \frac{1}{4}\sum^4_{i=1}\vec x_i,
\end{equation}
and the three relative coordinates $\{\vec X,\vec Y,\vec Z\}$ as
\begin{subequations}\label{relco}
\begin{eqnarray}
\label{XYdef}
    &&\vec X=\vec x_1-\vec x_2,\quad \vec Y=\vec x_3-\vec x_4, \\
\label{Zdef}
    &&\vec Z=\left[\frac{\vec x_1+\vec x_2}{2}\right]-\left[\frac{\vec x_3+\vec x_4}{2}\right].
\end{eqnarray}
\end{subequations}
The meaning of $\vec X$ and $\vec Y$ is obvious, while $\vec Z$ is the separation between the center of mass of the cluster made by gluons 1 and 2 and the corresponding one for gluons 3 and 4. A graphical representation of the four-gluon system and the relative coordinates is given in Fig.~\ref{fig3}.

We have shown in Sec.~\ref{4ganaly} that the internal structure of the $0^{+-}$ can be elucidated thanks to the helicity formalism provided that the particular permutation symmetry of this state is taken into account. A crucial point has now to be outlined: By definition of the relative variables $\vec X$ and $\vec Y$, the orbital angular momenta corresponding to clusters (1,2) and (3,4) are nothing else than $\vec L_X$ and $\vec L_Y$ respectively. Consequently, we are led to the following results
\begin{equation}\label{n1}
\left\langle \vec L^{\, 2}_X\right\rangle=\left\langle \vec L^{\, 2}_Y\right\rangle=4.
\end{equation}
These highly non standard values for a square orbital angular momentum are a direct consequence of the helicity formalism and could never have been deduced from an usual, spin-based, approach. A look at definition~(\ref{Zdef}) then shows that $\vec L_Z$ can be identified with the relative orbital angular momentum of the two clusters, so we obtain
\begin{equation}\label{n2}
\left\langle \vec L^{\, 2}_Z\right\rangle=0.
\end{equation}

\subsection{Numerical results}\label{rdiscu}
We have now all the necessary ingredients to compute at least a crude approximation of the mass of the lightest $0^{+-}$ state within our model. The mass that we will obtain can be compared to the lattice QCD data
\begin{equation}\label{data}
    r_0\, M_{0^{+-}}=11.54\pm0.24~\textrm{\cite{lat1}},\quad r_0\, M_{0^{+-}}=11.66\pm0.19~\textrm{\cite{lat3}}.
\end{equation}
Both results are actually compatible, the second one coming from a more recent computation.

The glueball mass we look for should \textit{a priori} be numerically computed from Hamiltonian~(\ref{mainH}). A full four-body calculation with such a semirelativistic Hamiltonian is a very difficult task in itself, but here the situation is even worse because the helicity formalism imposes non standard values for the matrix elements of the orbital angular momentum. Since the four-body computations are widely out of the current numerical methods at our disposal, we rather choose a ``digluon-digluon" approach: We will first compute the mass of the (1,2) and (3,4) clusters, and then calculate the mass of the bound state made of these two clusters. This scheme is very useful because only two-body calculations are needed, but also because we have investigated the internal structure of the $0^{+-}$ state in this particular scheme throughout this paper. Both gluon clusters should actually have the same mass $r_0\, M_c$, given by the lowest-lying state of Hamiltonian
\begin{equation}\label{hc}
r_0\, H_c=2\sqrt{\vec q\,^2_X}+X,
\end{equation}
where $\vec q_X$ and $\vec X$ are conjugate variables and where $X=|\vec X|$, with $\left\langle \vec L^{\, 2}_X\right\rangle=4$ from Eq.~(\ref{n1}). Such a Hamiltonian can be readily understood from Fig.~\ref{fig3}. Equation~(\ref{hc}) actually refers to the (1,2) cluster, but Hamiltonian $2\sqrt{\vec q^{\, 2}_Y}+Y$, valid for the (3,4) cluster, would lead to the same mass by symmetry. The eigenvalues of Hamiltonian~(\ref{hc}) can be accurately computed thanks to the Lagrange mesh method \cite{sem01}; we find $r_0\, M_c=4.72$.

As the masses of the clusters are known, the total mass $r_0\, M$ of the system can be computed. It should be given by the ground state of Hamiltonian
\begin{equation}
    r_0\, H_g=2\sqrt{\vec q^{\, 2}_Z+r_0^2M_c^2}+2Z.
\end{equation}
One can indeed conclude from Fig.~\ref{fig3} that the two ``digluons" should feel a confining potential such as if they were bound by two fundamental flux tubes. Applying again the Lagrange mesh method for $\left\langle \vec L^{\, 2}_Z\right\rangle=0$ [see Eq.~(\ref{n2})], one has
\begin{equation}
    r_0 M=11.61,
\end{equation}
in nice agreement with the lattice QCD data~(\ref{data}).

\section{Conclusions}\label{conc}

In the present work, we have given arguments showing that the glueballs and gluelumps that are currently observed in lattice QCD can be understood in terms of bound states of a few massless constituent gluons with helicity-1. In this scheme, the lowest-lying $C=+$ glueballs can be identified with two-gluon states (or at least with hadrons in which a two-gluon component widely dominates), while the lightest $C=-$ glueballs are compatible with three-gluon states and the heavy $0^{+-}$ glueball should be seen as a light four-gluon state. For what concerns gluelumps, we have shown that the lightest ones, which are $C=-$ states, are compatible with one-gluon states. This is allowed because of the static color octet source. To confirm this classification, we have shown that a linear behavior of the masses with respect to the expected number of constituent gluons is observed, suggesting that each constituent gluon brings a given energy to the total mass. The slope of the mass formula can be reproduced within a simple flux tube model with linear confinement and one-gluon-exchange potential provided that the gluons helicity is correctly taken into account. A useful concept has been introduced in order to classify the various gluonic states, that is the natural parity, defined by $(-1)^{N_g+J}$. It appears that a glueball with natural parity will always be lighter than the corresponding glueball with the same $J$ and $C$ but the opposite parity.

It is worth recalling that, if constituent gluons were regarded as spin-1 particles, some states would appear that are not observed in lattice QCD. The most striking ones would be light $1^{\pm+}$ glueballs in the two-gluon sector, but also a $0^{--}$ gluelump and a light $0^{--}$ three-gluon glueball. The absence of these states clearly favors helicity-1 constituent gluons.


In order to convince ourselves that the $0^{+-}$ glueball is a relevant four-gluon candidate, we have computed its mass starting from a four-body flux tube Hamiltonian. Due to the unusual color wave function and permutation symmetry of this glueball, only the confining potential is present and other relativistic corrections vanish. Other mechanism as instanton-induced interactions have been neglected. Moreover, the assumption that the constituent gluons are helicity-1 particles leads to the knowledge of the internal structure of the considered glueball. Once all these ingredients are put together, the mass we obtain is in agreement with lattice QCD without fitting any parameter. This brings a good support to the constituent-gluon picture we developed in this work.

The above results are actually a strong motivation to reconsider the various potential models of hadrons with a gluonic content under the assumption that constituent gluons are massless particles with a helicity-1. If two-gluon glueballs have already been studied in Ref.~\cite{gluh1}, the mass spectra of gluelumps, hybrid mesons, and three-gluon glueballs should now be explicitly computed within the full helicity formalism. Such a work is in progress.

\acknowledgments
F. B. and C. S. thank the F.R.S.-FNRS for financial support. V. M. thanks the IISN for financial support. The work of N.B. is supported in part by the EU contracts MRTN-CT-2004-503369 and MRTN-CT-2004-512194 and by the NATO grant PST.CLG.978785. The authors thank J. Nuyts for valuable discussions.

\begin{appendix}

\section{Helicity states for gluelumps}\label{hsgl}

Using the formalism developed in Ref.~\cite{gluh1}, it is easy to build the helicity states for a gluelump, that is a bound state with a gluon and a static $0^{++}$ color octet source, with $J \ge 1$,
\begin{eqnarray}
\label{hsgl1}
|1,0; J^P,M,1\rangle &=& \sqrt{\frac{J+1}{2J+1}}\ | ^3 J-1 _J \rangle +
\sqrt{\frac{J}{2J+1}}\ | ^3 J+1 _J \rangle \quad \textrm{with} \quad P=(-)^J, \\
\label{hsgl2}
|1,0; J^P,M,-1\rangle &=& -| ^3 J _J \rangle \quad \textrm{with} \quad P=(-)^{J+1}.
\end{eqnarray}
With these decompositions, a simple computation gives $\left\langle \vec L^{\, 2} \right\rangle=J(J+1)$.

\section{$C$-parity of few-gluon glueballs}\label{cval}

The gluonic field is defined as $A_\mu=A^a_\mu\, T_a$, where $T_a=\lambda_a/2$ are the generators of $\mathfrak{su}(3)$ and where $\lambda_a$ are the Gell-Mann matrices. If $\hat C$ is the $C$-parity operator, the charge conjugate of $A_\mu$, that we write $\tilde A_\mu$, is such that \cite{mink}
\begin{equation}\label{cconj}
    \tilde A_\mu=\hat C\, A_\mu\, \hat C^{-1}=-A_\mu^T.
\end{equation}
The index $^T$ denotes a transposition. Let us begin with the simple case of two-gluon states. The only way to make a color singlet is to write such a state as $\delta_{ab}\, A^a_\mu\, A^b_\nu$, $\delta_{ab}$ being the Kronecker symbol. As ${\rm Tr}(\lambda_a\lambda_b)=2\delta_{ab}$ \cite{grei}, we have $\delta_{ab}\, A^a_\mu\, A^b_\nu\propto {\rm Tr}(A_\mu A_\nu)$. It is then readily checked that ${\rm Tr}(\tilde A_\mu \tilde A_\nu)={\rm Tr}(A_\mu A_\nu)$, implying that two-gluon states always have a positive $C$-parity. Next, there are two ways of making a color singlet with three gluons: $f_{abc}A^a_\mu A^b_\nu A^c_\rho\propto{\rm Tr}(\left[A_\mu,A_\nu\right]A_\rho)$ and $d_{abc}A^a_\mu A^b_\nu A^c_\rho\propto{\rm Tr}(\{A_\mu,A_\nu\}A_\rho)$, with $f_{abc}$ the totally antisymmetric structure constant of $\mathfrak{su}(3)$ and $d_{abc}$ the totally symmetric ones. These are actually the $[\left[\bm 8,\bm 8\right]^{\bm 8_a},\bm 8]^{\bm 1_a}$ and $[\left[\bm 8,\bm 8\right]^{\bm 8_s},\bm 8]^{\bm 1_s}$ configurations respectively. It can then be computed that
\begin{equation}
    {\rm Tr}([\tilde A_\mu,\tilde A_\nu]\tilde A_\rho)={\rm Tr}(\left[A_\mu,A_\nu\right]A_\rho)\Rightarrow C=+,
\end{equation}
 \begin{equation}
    {\rm Tr}(\{\tilde A_\mu,\tilde A_\nu\}\tilde A_\rho)=-{\rm Tr}(\{A_\mu,A_\nu\}A_\rho)\Rightarrow C=-.
\end{equation}
These configurations have the permutation symmetries $[\mu\nu\rho]$ and $(\mu\nu\rho)$ respectively.

The case of four gluons is a little bit more involved.
Let us first focus on the particular color wave functions in which
the gluon pairs are in a color octet.
The following combinations lead to a color singlet:
$f_{abc}\, f^c_{\ de}\, A^a_\mu A^b_\nu A^d_\rho A^e_\sigma\propto {\rm Tr}([A_\mu,A_\nu][A_\rho,A_\sigma])$, $d_{abc}\, d^c_{\ de}\, A^a_\mu A^b_\nu A^d_\rho A^e_\sigma\propto \frac{1}{8}{\rm Tr}(\{\{ A_\mu,A_\nu\},A_\rho\}A_\sigma)-\frac{1}{6}{\rm Tr}(A_\mu A_\nu){\rm Tr}(A_\rho A_\sigma)$, and $f_{abc}\, d^c_{\ de}\, A^a_\mu A^b_\nu A^d_\rho A^e_\sigma\propto {\rm Tr}([A_\mu,A_\nu]\{A_\rho,A_\sigma\})$. They correspond respectively to the color wave functions  $[[\bm 8,\bm 8]^{\bm 8_a},[\bm 8,\bm 8]^{\bm 8_a}]^{\bm 1}$, $[[\bm 8,\bm 8]^{\bm 8_a},[\bm 8,\bm 8]^{\bm 8_s}]^{\bm 1}$, and $[[\bm 8,\bm 8]^{\bm 8_s},[\bm 8,\bm 8]^{\bm 8_s}]^{\bm 1}$. As previously, the $C$-parity can be computed. First,
\begin{equation}
    {\rm Tr}([\tilde A_\mu,\tilde A_\nu][\tilde A_\rho,\tilde A_\sigma])={\rm Tr}([A_\mu,A_\nu][A_\rho,A_\sigma])\Rightarrow C=+,
\end{equation}
with a permutation symmetry that can be of the form  $M^{|2,2|}_{\mu\nu\rho\sigma}\,$ only,
the total antisymmetry $[\mu\nu\rho\sigma]$ being excluded
by virtue of the Jacobi identity.
Second, we have
\begin{eqnarray}
    \frac{1}{8}{\rm Tr}(\{\{ \tilde A_\mu,\tilde A_\nu\},\tilde A_\rho\}\tilde A_\sigma)-\frac{1}{6}{\rm Tr}(\tilde A_\mu \tilde A_\nu){\rm Tr}(\tilde A_\rho \tilde A_\sigma)&=&\frac{1}{8}{\rm Tr}(\{\{ A_\mu,A_\nu\},A_\rho\}A_\sigma)-\frac{1}{6}{\rm Tr}(A_\mu A_\nu){\rm Tr}(A_\rho A_\sigma)\nonumber\\
    &\Rightarrow& C=+,
\end{eqnarray}
with a permutation symmetry that can be of the form $(\mu\nu\rho\sigma)$ or $M^{|2,2|}_{\mu\nu\rho\sigma}$. The only configuration that can lead to a negative $C$-parity actually is
\begin{equation}
    {\rm Tr}([\tilde A_\mu,\tilde A_\nu]\{\tilde A_\rho,\tilde A_\sigma\})=-{\rm Tr}([A_\mu,A_\nu]\{A_\rho,A_\sigma\})\Rightarrow C=-.
\end{equation}
\textit{A priori}, both the permutation symmetries $M^{|3,1|}_{\mu\nu\rho\sigma}$
and $M^{|2,1,1|}_{\mu\nu\rho\sigma}$ are allowed, but the equality
$f^{\ \ \ c}_{a(b}d_{de)c}=0$ forbids the $M^{|3,1|}_{\mu\nu\rho\sigma}$
symmetry. The last equality just reflects the obvious fact that,
for any three given matrices $A\,$, $B\,$ and $C\,$,
the following identity holds
$[A,\{B,C\}]+[B,\{A,C\}]+[C,\{A,B\}]= 0$.

The two remaining four-gluon color wave functions are now $[[\bm 8,\bm 8]^{\overline{\bm{10}}_a},
[\bm 8,\bm 8]^{\bm{ 10}_a}]^{\bm{1}}$ and
$[[\bm 8,\bm 8]^{\bm{27}_s},[\bm 8,\bm 8]^{\bm{ 27}_s}]^{\bm{1}}$. The study of these configurations can be achieved by building the various representations we need  from bilinear
quantities in the fundamental irreducible representation of $\mathfrak{su}(3)$
and its conjugated, $\bm{3}$ and $\overline{\bm{3}}\,$,
using $\bm{3}\otimes\overline{\bm{3}}= \bm{8} \oplus \bm{1}\,$ \cite{carru}. Let us take a complex 3-vector $V_j$ and its complex conjugated $\bar{V}^k$ transforming in the
$\bm{3}$ and $\overline{\bm{3}}$ representations respectively, where $j,k = 1,2,3\,$,
and impose the vanishing of the trace $\bar{V}^l V_l=0\,$.
Then, $T^j_{\;k}=i\,\bar{V}^j V_k\,$ defines an anti-hermitian traceless $3\times 3$ matrix, therefore transforming in the adjoint irrep of $\mathfrak{su}(3)$, $(T^j_{\;k})^{*}=-i\,(\bar{V}^j V_k)^{*}=-i\,(\bar{V}^k V_j)=$
$-T^k_{\;j}\,$. This relation is actually related to Eq.~(\ref{cconj}). Note that the dual irreps of $\mathfrak{su}(N)$ are
equivalent to the complex conjugated ones, implying for
$\mathfrak{su}(3)\,$ that
\begin{equation}\label{uab}
\bar{U}^k_{^{[A,B]}} \simeq A_i\, B_j \,\varepsilon^{ijk}
\end{equation}
belongs to the $\bar{\bm{3}}$ representation, the symbol $\simeq$ meaning `isomorphic to'. The tensor product $\bm{ 8}\otimes\bm{8}$ can then be performed explicitly, giving the well-known result
\begin{equation}\label{decomp2}
\bm{8}  \otimes \bm{8}=\bm{27}_s  \oplus \bm{\overline{10}}_a
\oplus \bm{10}_a \oplus \bm{8}_a  \oplus \bm{8}_s \oplus \bm{1}_s,
\end{equation}
where the permutation symmetry, i.e. symmetric ($s$) or antisymmetric ($a$), of all the representations have explicitly been written.
Starting from $(\bar{A}^iA_j) (\bar{B}^kB_l) $, the left-hand-side
of the above relation, we find the correct decomposition with
\begin{eqnarray}
\bm{1}_s  & \simeq & Z(A,B) \equiv (\bar{A}^p {A}_q)
 (\bar{B}^q {B}_p)\;, \quad \bm{ 8}_s \simeq  \widehat{Y}^i_{\;j}(A,B) \equiv
(\bar{A}^i {A}_p) (\bar{B}^p {B}_j)+
(\bar{B}^i {B}_p) (\bar{A}^p {A}_j)
-\frac{2}{3}\;\delta^i_j\,Z(A,B) \;,
\nonumber \\
\bm{ 8 }_a &\simeq &
(\bar{A}^iA_k) (\bar{B}^k B_j) - (\bar{B}^iB_k) (\bar{A}^k A_j ) \;,
\quad
\bm{ \overline{10} }_a \simeq
\bar{A}^{(i}\bar{B}^{k}
\bar{U}^{m)}_{^{[A,B]}} \;,
\quad
\bm{ 10}_a \simeq
{U}_{(m}^{_{[A,B]}}{A}_{j}{B}_{l)} \;,
\nonumber \\
\bm{ 27}_s &\simeq & \widehat{X}^{ik}_{jl}(A,B)\equiv
X^{ik}_{jl}(A,B) - \frac{1}{5}\, \delta^{(i}_{(j}
\widehat{Y}^{k)}_{\;l)}(A,B)
 - \frac{1}{12}\, \delta^{(i}_{(j} \delta^{k)}_{l)}\,Z(A,B)\;,
\nonumber
\end{eqnarray}
with $X^{ik}_{jl}(A,B) = \bar{A}^{(i}\bar{B}^{k)} {A}_{(j} {B}_{l)}$
and $\widehat{X}^{ik}_{jk}(A,B)=0$.
The action of the $C$-parity on all these irreducible
representations is particularly simple: It lowers (raises) the upper (lower)
indices and removes (adds) bars accordingly.
For example, $A_j$ is changed into $\bar A^j$. It is readily
checked that the $\bm{1}_s$ has a positive $C$-parity, and that all
the previous results concerning four gluons are recovered. In the case of a
four-gluon system, the decomposition~(\ref{decomp2}) tells us that two singlets
can be obtained for gluon pairs in $\bm{10}_a$ and $\overline{\bm{10}}_a$.
These are given by
\begin{eqnarray}
    \bm{1}_\pm \simeq U^{[A,B]}_{(m}A_j B_{l)}
\bar{U}^{(m}_{[C,D]}\bar C^j\bar{D}^{l)}\pm \bar{U}_{[A,B]}^{(m}
\bar{A}^j \bar{B}^{l)} U_{(m}^{[C,D]} C_j D_{l)}.
\end{eqnarray}
This last line can be used to check that the $\bm{1}_+$ ($\bm{1}_-$) singlet has a positive (negative) $C$-parity. Finally, their permutation symmetry is of $M^{|2,2|}$ type. The configurations involving the $\bm{27}_s$ representation generate repulsive forces between the gluons. They will necessarily lead to heavy glueballs and will consequently not be studied here, were attention is drawn on low-lying states.

A few comments can be done concerning gluelumps. The color octet source can actually be seen as a color scalar particle $\phi=\phi^a\, T_a$ such that $\tilde \phi=\phi^T$. Then, a one-gluon gluelump corresponds to ${\rm Tr}(\phi A_\mu)$, and it is easily checked that the $C$-parity of such a state is always negative. For two-gluon gluelumps, one can further compute that $C=+$ ($-$) for a totally symmetrical (antisymmetrical) color configuration.

\end{appendix}


\begin{thebibliography}{99}
\bibitem{exp_gg} E.~Klempt and A.~Zaitsev, Phys. Rept. {\bf 454}, 1 (2007) [arXiv:0708.4016], and references therein.
\bibitem{lat1} C. J. Morningstar and M. Peardon, Phys. Rev. D \textbf{60}, 034509 (1999) [hep-lat/9901004].
\bibitem{lat2} H. B. Meyer and M. J. Teper, Phys. Lett. B \textbf{605}, 344 (2005) [hep-ph/0409183].
\bibitem{lat3} Y. Chen \textit{et al.}, Phys. Rev. D \textbf{73}, 014516 (2006) [hep-lat/0510074].
\bibitem{cg} A.~P.~Szczepaniak and E.~S.~Swanson, Phys. Lett. B {\bf 577}, 61 (2003) [hep-ph/0308268].
\bibitem{bar} T. Barnes, Z. Phys. C \textbf{10}, 275 (1981); A. B. Kaidalov and Yu. A. Simonov, Phys. Lett. B \textbf{636}, 101 (2006) [hep-ph/0512151].
\bibitem{bar2} J.~M.~Cornwall and A.~Soni, Phys. Lett. B {\bf 120}, 431 (1983).
\bibitem{brau} F.~Brau and C.~Semay, Phys. Rev. D {\bf 70}, 014017 (2004) [hep-ph/0412173].
\bibitem{gvm} V.~Mathieu, C.~Semay, and B.~Silvestre-Brac, Phys.\ Rev.\ D {\bf 74}, 054002 (2006) [hep-ph/0605205].
\bibitem{gvm2} V.~Mathieu, C.~Semay, and B.~Silvestre-Brac, Phys. Rev. D \textbf{77}, 094009 (2008)  [arXiv:0803.0815].
\bibitem{gluh1} V. Mathieu, F. Buisseret, and C. Semay, Phys. Rev. D {\bf 77}, 114022 (2008) [arXiv:0802.0088].
\bibitem{gl2} M. Foster and C. Michael, Phys. Rev. D \textbf{59}, 094509 (1999) [hep-lat/9811010].
\bibitem{gl1} G. S. Bali and A. Pineda, Phys. Rev. D \textbf{69}, 094001 (2004) [hep-ph/0310130].
\bibitem{mata} An complete review and many references about baryons in large-$N_c$ QCD can be found in N. Matagne, PhD thesis, University of Li\`ege, 2006 [hep-ph/0701061].
\bibitem{witten} E. Witten, Nucl. Phys. B \textbf{160}, 57 (1979).
\bibitem{lnc} C. Semay, F. Buisseret, N. Matagne, and Fl. Stancu, Phys. Rev. D \textbf{75}, 096001 (2007) [hep-ph/0702075]; C. Semay, F. Buisseret, and Fl. Stancu, Phys. Rev. D \textbf{76}, 116005 (2007) [arXiv:0708.3291].
\bibitem{liu} C.~Liu, Eur.\ Phys.\ J.\  C {\bf 53}, 413 (2008) [arXiv:0710.4185].
\bibitem{cg0} A.~P.~Szczepaniak and E.~S.~Swanson, Phys.\ Rev.\  D {\bf 65}, 025012 (2002) [hep-ph/0107078].
\bibitem{cg1} A.~P.~Szczepaniak and E.~S.~Swanson, Phys.\ Rev.\ Lett.\  {\bf 87}, 072001 (2001) [hep-ph/0006306].
\bibitem{cg2} A.~Szczepaniak, E.~S.~Swanson, C.~R.~Ji and S.~R.~Cotanch, Phys.\ Rev.\ Lett.\  {\bf 76}, 2011 (1996) [hep-ph/9511422]; A.~P.~Szczepaniak and E.~S.~Swanson, Phys.\ Lett.\  B {\bf 577}, 61 (2003) [hep-ph/0308268].
\bibitem{llan1} F.~J.~Llanes-Estrada, P.~Bicudo, and S.~R.~Cotanch, Phys.\ Rev.\ Lett.\  {\bf 96}, 081601 (2006) [hep-ph/0507205].
\bibitem{first} T. Barnes, Z. Phys. C \textbf{10}, 275 (1981); A. B. Kaidalov and Yu. A. Simonov, Phys. Lett. B \textbf
{636}, 101 (2006) [hep-ph/0512151]; F. Brau and C. Semay, Phys. Rev. D \textbf{70}, 014017 (2004) [hep-ph/0412173].
\bibitem{glulat} C.~J.~Morningstar and M.~J.~Peardon, Phys.\ Rev.\  D {\bf 56}, 4043 (1997) [hep-lat/9704011].
\bibitem{book} J. Smit, \textit{Introduction to Quantum Fields on a Lattice} (Cambridge University Press, 2002).
\bibitem{jaffe} R.~L.~Jaffe, K.~Johnson and Z.~Ryzak, Annals Phys.\  {\bf 168}, 344 (1986).
\bibitem{effpot} F. Buisseret and C. Semay, Eur. Phys. J. A \textbf{33}, 87 (2007) [hep-ph/0611216].
\bibitem{jaco} M. Jacob and G. C. Wick, Ann. Phys. \textbf{7}, 404 (1959).
\bibitem{gie} D. R. Giebink, Phys. Rev. C \textbf{32}, 502 (1985).
\bibitem{wick3} G. C. Wick, Ann. Phys. \textbf{18}, 65 (1962).
\bibitem{gross} A. Stadler, F. Gross, and M. Frank, Phys. Rev. C \textbf{56}, 2396 (1997).
\bibitem{tung} W.~K.~Tung, \textit{Group Theory In Physics} (World Scientific, Singapore, 1985).
\bibitem{fulh} W. Fulton and J. Harris, \textit{Representation Theory: A First Course} (Springer-Verlag, New-York, 1991).
\bibitem{glusp} P.~O.~Hess \textit{et al.}, Eur.\ Phys.\ J.\  C {\bf 9}, 121 (1999) [hep-ph/9810404].
\bibitem{yang} C. N. Yang, Phys. Rev. \textbf{77}, 242 (1950).
\bibitem{mink} H. Fritzsch and P. Minkowski, Nuovo Cimento \textbf{30} A, 393 (1975).
\bibitem{fumi} F. G. Fumi and L. Wolfenstein, Phys. Rev. \textbf{90}, 498 (1953), and references therein.
\bibitem{oge} V. Mathieu and F. Buisseret, J. Phys. G \textbf{35}, 025006 (2008) [hep-ph/0702226].
\bibitem{sczg} P.~Guo \textit{et al.}, Phys.\ Rev.\ D {\bf 77}, 056005 (2008) [arXiv:0707.3156].
\bibitem{sem01} C. Semay, D. Baye, M. Hesse, and B. Silvestre-Brac,\ Phys. Rev. E\ \textbf{64},\ 016703\ (2001).
\bibitem{model} P. Bicudo, M. Cardoso, and O. Oliveira, Phys. Rev. D \textbf{77}, 091504(R) (2008) [arXiv:0704.2156]; M. Cardoso and P. Bicudo, arXiv:0807.1621.
\bibitem{math} V. Mathieu, C. Semay, and F. Brau, Eur. Phys. J. A {\bf 27}, 225 (2006) [hep-ph/0511210].
\bibitem{grei} Useful properties of the Gell-Mann matrices can be found for example in W. Greiner and B. M\"{u}ller, \textit{Quantum mechanics: symmetries} (Springer, Berlin, 1994).
\bibitem{carru} P. Carruthers, \textit{Introduction to unitary symmetry}, (Interscience Publishers, USA,1966).

\end{thebibliography}
\end{document}